\documentclass[11pt]{article}
\usepackage[utf8]{inputenc}
\usepackage[english]{babel}
\usepackage[T1]{fontenc}
\usepackage{amsmath,amssymb,amsthm,graphicx}
\usepackage{color}
\usepackage[top=1.5in, bottom=1.8in]{geometry}
\usepackage[labelfont=sc]{caption}
\usepackage[colorlinks=true,citecolor=blue,linkcolor=black,urlcolor=blue]{hyperref}
\usepackage{siunitx}

\usepackage{microtype}
\usepackage{lmodern} 

\newcommand{\citep}{\cite}
\newcommand{\tc}[1]{#1} 

\title{Statistical modelling of conidial discharge of entomophthoralean fungi using a newly discovered \speccite{Pandora} species}
\author{Niels Lundtorp Olsen\textsuperscript{1}, Pascal Herren\textsuperscript{2}, \\  Bo Markussen\textsuperscript{1}, Annette Bruun Jensen\textsuperscript{2}, Jørgen Eilenberg\textsuperscript{2}}
\date{April 6, 2019}

\newcommand{\pil}{\rightarrow}

\newcommand{\R}{\mathbb{R}}

\newcommand{\N}{\mathbb{N}}

\newcommand{\bu}{\boldsymbol{u}}
\newcommand{\bgamma}{\boldsymbol{\gamma}}
\newcommand{\E}{\mathrm{E}}

\newcommand{\de}{\: \mathrm{d}}
\newcommand{\inte}[4]{\int_{#1}^{#2} \! #3 \de \mathrm{#4} }

\newcommand{\trans}{^\top \!}
\newcommand{\gradc}{{$^\circ$C}{}}

\newcommand{\speccite}[1]{\textit{#1}}
\newcommand{\figmark}{Figure~}

\theoremstyle{definition}

\newcommand{\linjeafstand}{1.05}

\begin{document}

\vfill	
\def\spacingset#1{\renewcommand{\baselinestretch}%
	{#1}\small\normalsize} \spacingset{1}
\maketitle


\begin{abstract}
	Entomophthoralean fungi are insect pathogenic fungi and are characterized by their active discharge of infective conidia that infect insects.
	
	Our aim was to study the effects of temperature on the discharge and to characterize the variation in
	the associated temporal pattern of a newly discovered \speccite{Pandora} species with focus on peak location and shape of the discharge.
	
	Mycelia were incubated at various temperatures in darkness, and conidial discharge was measured over time.
We used a novel modification of a statistical model (pavpop), that simultaneously estimates phase
and amplitude effects, into a setting of generalized linear models. 	
This model is used to test hypotheses of peak location and discharge of conidia. The statistical analysis showed that high temperature leads to an early and fast decreasing peak, whereas there were no significant differences in total number of discharged conidia. Using the proposed model we also quantified the biological variation in the timing of the peak location at a fixed temperature.

\end{abstract}	

\vfill

\textbf{1} Department of Mathematical Sciences, \textit{University of Copenhagen}
\\
\textbf{2} Department of Plant and Environmental Sciences, \textit{University of Copenhagen}

\thispagestyle{empty}

\newpage
	
\spacingset{\linjeafstand}	
	
\section{Introduction}

\tc{Fungi are important as biological control agents, and their effect is due to their infective spores \cite{hajekeilenberg}. The mechanisms for spore releases differ among fungal taxons, one mechanism is shooting off spores as found in fungus order Entomophthorales, which are insect and mite pathogens.  
Conidia are the infective units of entomophthoralean fungi, and for the majority of species they
are actively discharged \citep{shah2003entomopathogenic}.	The large conidia (mostly between 15 and 40 microns in length) in {Entomophthorales} demands high energy to be discharged. The spore discharge mechanism for entomophthoralean fungi allow fungi to convert elastic energy into kinetic energy, ensuring that spores are discharged at sufficient speeds. }

\begin{figure}[!htb]
\centering { \includegraphics[width=0.5\textwidth]{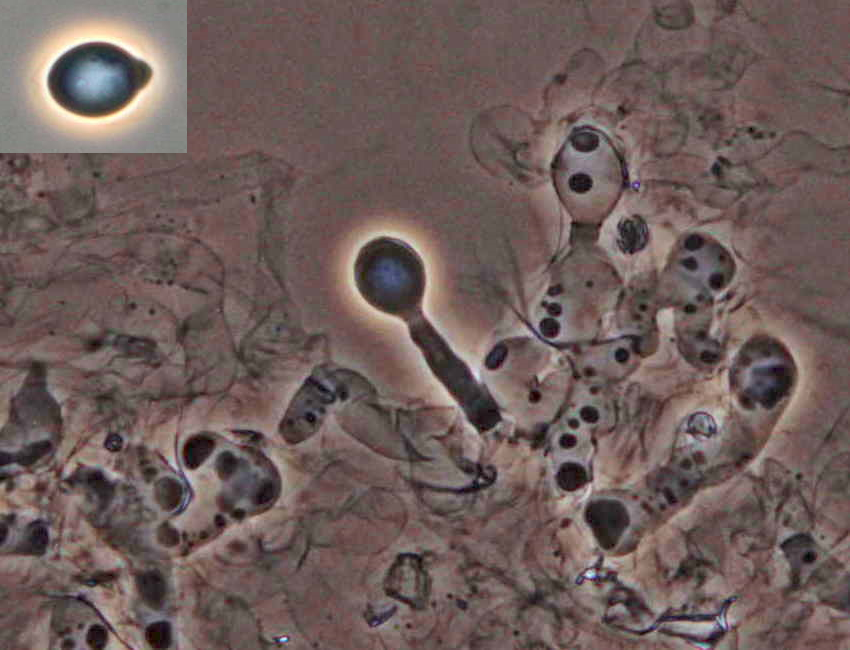}} \caption{The mycelium of \speccite{Pandora} sp. from \speccite{Cacopsylla} sp. with a primary conidium on top of a conidiophore and a discharged primary conidium (insert in upper left corner). The length of conidia of \speccite{Pandora} sp. is  15.6-23.3 \si{\micro\metre} \cite{ahjensen}.
	} \label{fig-conidia-picture}
\end{figure}

The infection success depends, among other things, on the attachment of the discharged conidium after landing on host cuticle \citep{boomsma2014}. The conidia of entomophthoralean fungi are discharged with fluid from the conidiophore, which further assist the conidium to stick to host cuticle after landing \citep{money2016, eilenberg1987}.  Once the conidia of entomophthoralean fungi are discharged from the conidiophores they have a short longevity \citep{hajekmeyling, furlong1997}. 
\figmark \ref{fig-conidia-picture} shows a conidium of \speccite{Pandora} sp., a species from Entomophthorales isolated from an infected Psyllid in 2016 \cite{ahjensen}.

The temporal pattern of conidial discharge from infected and dead hosts  have been studied for several species of Entomophthorales belonging to the genera \speccite{Entomophthora}, \speccite{Entomophaga}, \speccite{Pandora} and \speccite{Zoophthora}
\citep{eilenberg1987, aoki1981, hemmati2001conidial, hajek2002andothers,wraight2003}. The studies show the same overall pattern: after a lag phase of a few hours after the death of the host, conidial discharge is initiated. Depending on host species, fungus species, and temperature, the peak in discharge intensity will be reached within one or two days, thereafter the intensity drops although conidia may still be produced and discharged several days after death of the host.  In principle the same pattern appears when conidia are discharged from in vitro cultures. Here the starting point will be when a mycelium mat{, which has been grown on nutritious solid medium,} is transferred onto for example moist filter paper or water agar, from where conidial discharge will be initiated. 
{ The conidial discharge of different species of Entomophthorales is affected by temperature {in relation to the intensity and the total number of discharged conidia} \citep{yu1995studies,hemmati2001conidial, shah2002effects, li2006factors}. }

\tc{It is essential to study mechanisms of spore discharge in insect pathogenic fungi \cite{eilenberg1986} , effects of environmental factors on spore discharge and also, modelling of spore dispersal in the field \cite{hesketh}.
	As pointed out by \cite{hesketh}, there is a need for more studies on spore discharge and the modelling of dispersal in order to understand natural ecosystem functioning and in order to develop more biological control based on fungi.
}
It is, \tc{however,} a methodological challenge to study patterns at a quantitative level over time of conidial discharge in Entomophthorales since the system is very dynamic and conidia are sticky. People have therefore used various methods to collect and count discharged conidia. In \cite{hajek2012} different methodologies applied to entomophthoralean fungi are reviewed, and a common trait is that the setup should as much as possible reflect the natural condition, where insects are killed and thereafter initiate discharge of conidia. Different laboratory setups have been used for obtaining discharged conidia counted on glass slides referring to specific time intervals and/or different distances \citep{eilenberg1987, hemmati2001conidial, kalsbeek2001}.  The data treatments in studies on conidial discharge are mostly rather simple and include for example calculations of mean and standard deviation for replicates, pairwise comparisons or analysis of variance, and a description in words about peak of intensity and length of period with conidial discharge. While these methods are valid and may offer a fair background for conclusions, they nevertheless do not make use of the total {biological} information in the study.

\paragraph{Statistical modelling of  temporal variation in biological systems using functional data}
 For biological processes that progress over time, we may like to think in terms of idealised systems with a clear time-dependent profile. However, it is often the case that different instances/replications of such processes show some variation in timing. Within statistics, this variation is commonly referred to as \emph{temporal variation} or \emph{phase variation}.
	 
	 The usual interpretation of temporal variation is \emph{biological time}; that is, the clock of the underlying biological system is out of synchronization with the idealised system (this may be for various reasons), but it is the same underlying processes that are taking place. A common example is puberty for boys: healthy boys enter the pubertal stage (which has some common characteristics for all boys) at some point, but when that happens more exactly varies \tc{considerably} between individuals.
	 
	 Inferring the effect of biological time obviously requires replications of the same experiment, and when the underlying structure is a continuous process, such data {is} naturally handled within the framework of functional data analysis (FDA) \citep{ramsay1982}. 
\tc{	The use of FDA will allow us to estimate sharply defined curves and to estimate and adjust for the variation in peak position \cite{ramsaybog}.
	 }
We believe that the effects of temporal variation are {sometimes} neglected in the biological sciences{, something which} can lead to weak or in worst cases even misleading conclusions \cite{ramsaybog}.

	 There are various approaches to modelling functional data with temporal variation (misaligned functional data).
	 We intend to follow the {novel} methodology of \cite{RaketSommerMarkussen}, which we will refer to as the \emph{pavpop} model (\textit{Phase and Amplitude Variation of POPulation means}). This methodology has been used in different applications with great success \citep{olsen2018, RaketGrimme2016}.
The main idea of  \cite{RaketSommerMarkussen} is simultaneous modelling of amplitude and temporal variation, where temporal variation is modelled as a spline interpolation of a latent Gaussian variable that represents temporal deviation from the idealised system. 
  For a review of methods for handling misaligned functional data, we refer to \citep{RaketSommerMarkussen, olsen2018}.
	  
	  Whereas classification is often part of papers on misaligned functional data, inference in form of hypothesis testing has got little attention in misaligned functional data. In general, inference in functional data is not easy and requires either strong parametric assumptions, which can be wrong, or the use of non-parametric tests, which can be computationally difficult.

\paragraph{Purpose and content of this study}
Overall, we aimed at getting a better understanding of the temporal progression of conidial discharge in entomophthoralean fungi  by applying dedicated methods from functional data analysis (FDA). 
\tc{ To the authors' knowledge, FDA has not been applied to this kind of studies before, this is a secondary aim of this study.}
The novelty of this work is also the extension and application of the pavpop model to discrete data, which are generated from an unobserved biological system with temporal progression. We  consider inferential questions,  which is new to this methodology as well.
In a more broad context, this can be seen as combining the pavpop model with generalized linear models. In this application we use a negative binomial response model; {the supplementary material describes various other response models. }
	
In this study,  discharge of \speccite{Pandora} conidia (\figmark \ref{fig-conidia-picture}) as a function of time was studied at different temperatures. {We hypothesize that 
	high temperature leads to an early and fast decreasing peak when looking at (the intensity of) conidial discharge, whereas a low temperature leads to a late and more slowly decreasing peak}, and  that a low temperature leads to a higher total production of conidia in the first 120 hours, as compared with higher temperatures.

\section{Methods}

\subsection{Experiment and data collection}

{A detailed description of the data collection can be found in \cite{herrenthesis}.} 
\medskip

We used an isolate of \textit{Pandora} sp. (KVL16-44). It is an insect pathogenic fungus found first time in 2016 in Denmark on  \speccite{Cacopsylla} sp. (Hemiptera). The conidia have typical \speccite{Pandora} morphology, including mononucleate conidia \cite{ahjensen}. The fungus is however a new, yet undescribed species based on molecular sequencing \cite{ahjensen}, and we will therefore in our paper use the name \speccite{Pandora} sp.

\paragraph{Mycelium production:}
The fungus was grown on Sabouraud Dextrose Agar (SDA) supplemented with egg yolk and milk \cite{hajek2012} in Petri dishes (55 mm diameter). To produce fresh material mycelium mats were transferred to new petri dishes  and incubated at 18.5 $^\circ$C in dark conditions for 20 days. 	
Using mycelial mats has the advantage, compared to use fungus killed insects, that conidia production can be syncronized more precisely.

\paragraph{Conidia production:}
Filter papers of 18 x 18 mm, moistened with 0.75 ml of autoclaved water, were placed in the center of lids of petri dishes (34 mm diameter). Four squares of 5 x 5 mm mycelium mats were cut from the same mycelium mat 20 mm away from the centre of the Petri dish and put upside down in the edges of the moist filter papers. All lids were put on the counter parts of the empty Petri dishes and they were kept in three different temperatures (12.0, 18.5 and 25.0 $^\circ$C) in complete darkness at 100\% RH. The mycelium mats were facing downwards. For each temperature, five replicates were made.  

\paragraph{Conidia discharge over time:}
To measure the conidial discharge a small stripe of Parafilm with a cover slip (18 x 18 mm) on top was placed inside the lower part of each petri dish. The dishes were then placed in  incubators (12.0, 18.5 and 25.0 $^\circ$C) for 30 min. The cover slips were placed underneath the four mycelium mats. The cover slips	 were removed immediately afterwards, and the conidia laying on the slip were stained with lactic acid (95\%). This procedure was repeated every eight hours for 120 h, which meant that in total, we obtained observations from 16 time points. The lower parts of the Petri dishes were cleaned with ethanol (70\%) and demineralised water every eight hours to ensure that primary conidia did not discharge secondary conidia on the cover slips. 
The conidia were counted in each of the four corners of the cover slips. In total, we got four observations per time-point, replicate and temperature (4 counts * 5 replicates * 16 time-points * 3 temperatures = 960 observations). Conidia on cover slips were counted with the aid of a light microscope (Olympus Provis) at x 400 magnification.

\subsection{Statistical modelling} \label{stat-afsnit}
	
	We consider a set of $N = 15$ latent \emph{mean curves}, $u_1, \dots, u_N : [0, 1] \pil \R$ 
	from $J = 3$ treatment groups. 
	The mean curves are assumed to be {independently} generated according to the following model
	\begin{equation}
	u_n(t) = \theta_{f(n)} (v_n(t)) + x_n(t), \quad n = 1, \dots, N \label{u-def}
	\end{equation}
	where $f$ maps curves into treatment groups.
	That is, to each subject corresponds a fixed effect $\theta_j$, which is perturbed in time by $v_n$ and in amplitude  by $x_n$, both assumed to be random. The temporal perturbation $v_n$ is usually referred to as a \emph{warping function}. 
	
	To each curve corresponds a set of discrete observations $(t_{n1}, y_{n1}), \dots, (t_{nm_n}, y_{nm_n}) \in [0,1] \times \mathcal{Y}$ where $(t_{n1}, \dots , t_{nm_n})$ are $m_n$ pre-specified time points and $\mathcal{Y} \subseteq \R$ is the sample space for the observations. 
	
	We assume that the observations conditionally on the latent mean curves are independently generated from an exponential family with probability density function 
\begin{equation}
	p(y | \eta) = b(y) \exp( \eta \cdot y - A( \eta, y)), \quad \eta \in \R, y \in \mathcal{Y} \label{eksp-familie}
\end{equation}
  where $\eta$ is the value of the latent mean curve at a given time, and $y$ is the canonical statistic for the observations. $A$ and $b$ are functions defining the exponential family. 
	We assume that $A(\eta, y)$ is two times continuous differentiable in $\eta$  with the property that $A''_\eta(\eta , y) > 0$ for all $\eta$ and $y$,
	and we assume that all hyperparameters describing $A$ and $b$ are known and fixed beforehand.
	More details on  response models can be found in the supplementary material. 
	
	The amplitude variation $x_n$ is assumed to be a zero-mean Gaussian process.
	The fixed effects $\theta_n$ are modelled using an appropriate spline basis, {and the warping functions $v_n$ are  parametrised  by Gaussian variables $w_n \in \R^{m_w}$ such that $w_n = 0$ corresponds to the identity function on $[0,1]$.} More details on fixed effects and phase variation can be found in the supplementary material. 
	
Estimation in this model is presented in the supplementary material. 	
Estimation is a major challenge, as direct estimation is not feasible due to the large number of latent variables.  Furthermore, unlike \cite{olsen2018}, the response is not Gaussian, which require additional considerations. We propose to use a twofold Laplace approximation for doing approximate maximum likelihood estimation; details on the Laplace approximation are found in the supplementary material.

\subsection{Data analysis}
As described in the data collection section, data consist of 960 observations (4 counts * 5 replicates * 16 time-points * 3 temperatures) in $\N_0$. The largest count was 211, and a large fraction of the counts was zero. 

Samples no. 7 and no. 13 effectively terminated the discharge of conidia after 48 and 40 hours, respectively, and therefore measurements from these samples were cancelled after these hours.

\paragraph{Response model}

A popular choice for modelling count data from biological experiments are Poisson models. This is backed by a strong theoretical reasoning; using our data as an example,  
one would expect that while the fungi are placed in the incubators, they would independently discharge conidia at random and at a constant rate. This would be a typical example of a Poisson process. 

\begin{figure}[!htb]
		\includegraphics[width=0.6\textwidth]{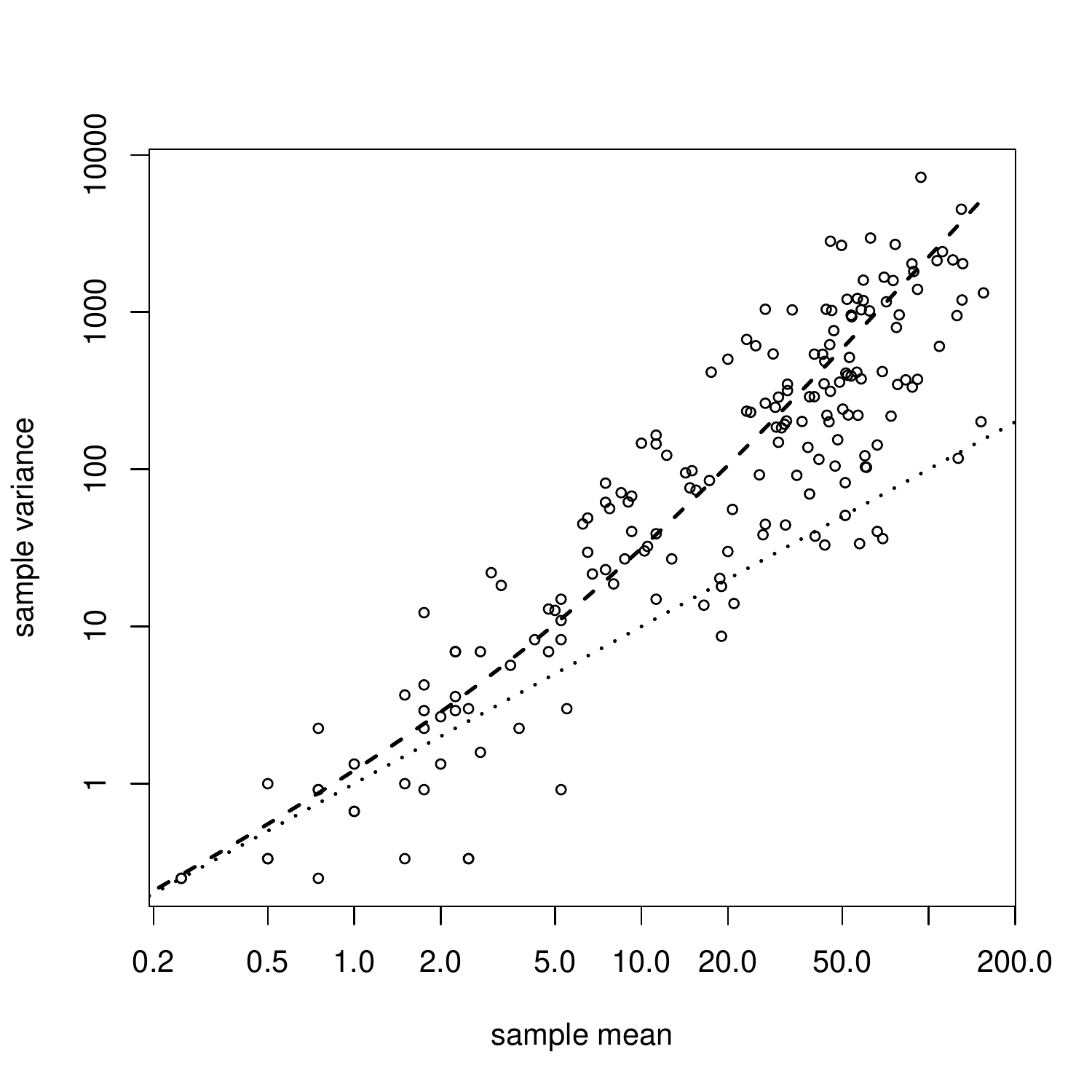} 
		\caption{Sample variance as a function of sample means across measurements. Dashed line is fit using a NB(4.66)-model; dotted line is fit using a Poisson model \tc{Each sample consisted of four observations.}}  \label{mean-var-nb}
\end{figure}

However, a unique feature of the data set was the four samples taken from each batch used for conidia count, which can reasonably be assumed to be independent and identically distributed conditionally on the latent curve $u$.
By comparing sample means and sample variances across the 240 measurements, this allowed us to assess if data was in reasonable agreement with  a Poisson model or overdispersed relative to this.
 As indicated in \figmark \ref{mean-var-nb}, data was clearly overdispersed; the Poisson model corresponds to the dotted line. 
 
Because data was overdispersed,  we instead fitted an unstructured negative binomial regression model with common rate $r$ to the data instead. 
This was in good concordance with data: the estimated rate was $r_0 = 4.658$, and the dashed line in \figmark \ref{mean-var-nb} indicates the corresponding mean/variance relation. 
	This value was fixed and used in the subsequent analysis. 
	
\begin{figure}[!htb]
	{\centering
	\includegraphics[width=0.95\textwidth]{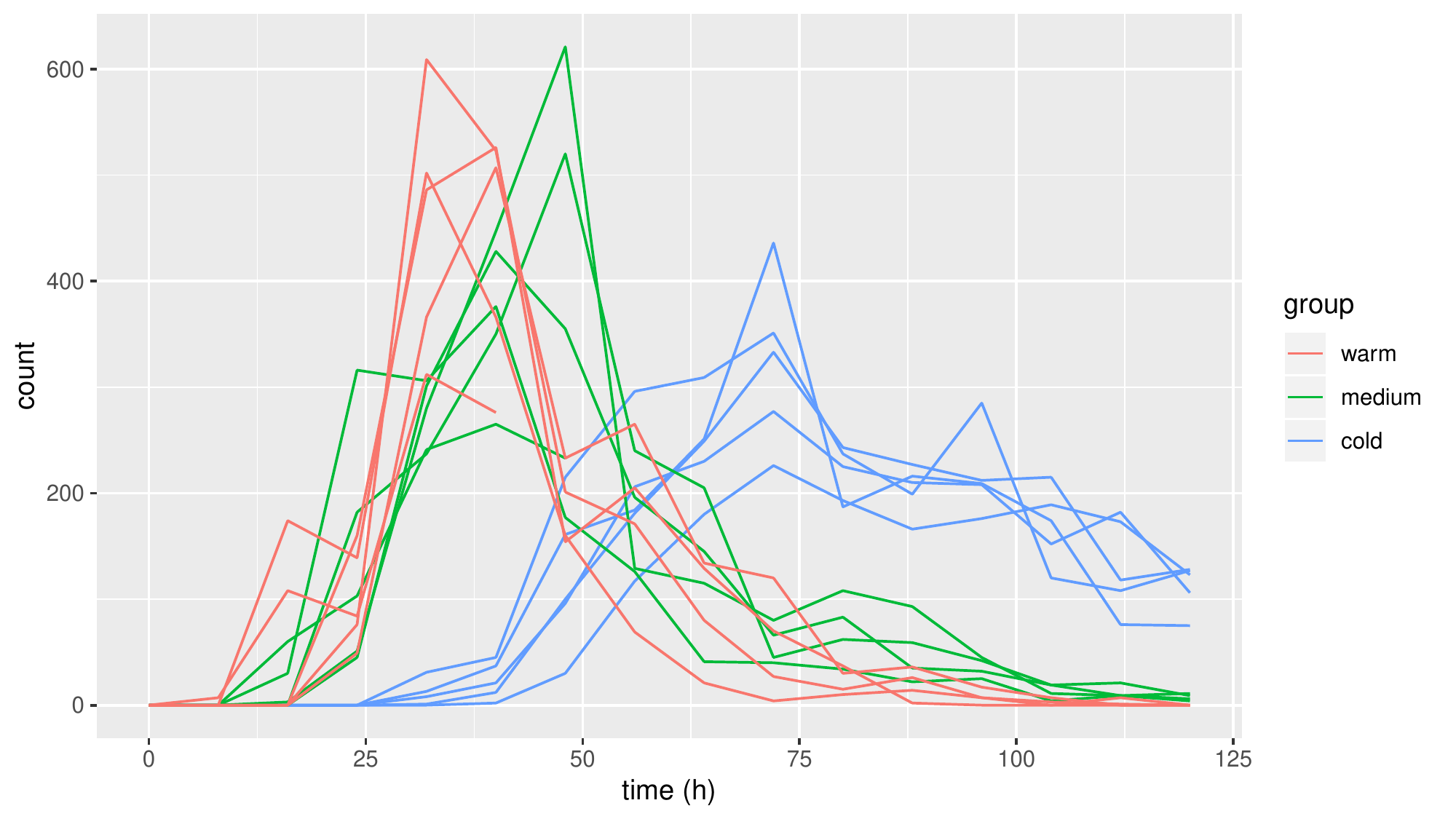}
	\caption{Summed counts of conidia for the individual fungi as functions of time, color-coded according to temperature.}  \label{fig-data-sum}}
\end{figure}

	Having estimated the dispersion, the counts at individual measurements were added for the subsequent analysis as the sum of counts is a sufficient statistic for our model. The sum of independent and identically distributed negative binomial random variables is again negatively binomially distributed; the rate parameter $r$ is multiplied by the number of counts; thus we got $r = 4 \cdot r_0 = 18.63$. The summed counts are displayed in \figmark \ref{fig-data-sum}.

\paragraph{Model for mean curves}
Time was rescaled to the unit interval such that $t=0$ corresponded to 0 hours and $t=1$ corresponded to 120 hours.
Warping functions were modelled as increasing cubic (Hyman~filtered) splines with $m_w = 7$
equidistant internal anchor points with extrapolation at the right boundary point. The latent variables $w_n$ were modelled using a Matérn covariance function with smoothness parameter $\alpha = 3/2$ and unknown range and scale parameters. This corresponds to discrete observations of an integrated Ornstein-Uhlenbeck process.
This gave a flexible, yet smooth, class of possible warping functions which also take into account that the internal clocks of individual fungi could be different at the end of the experiment.

Population means $\theta_{\text{cold}},\theta_{\text{medium}},  \theta_{\text{warm}}$ were modelled using natural cubic splines with 11 basis functions and equidistant knots in the interval $[0,1]$. Natural cubic splines are more regular near boundary points than b-splines which reduce the effect of warping on estimation of spline coefficients.

Amplitude covariance was modelled using a Matérn covariance function with unknown range, smoothness and scale parameters; see supplementary material for details.

\paragraph{Hypotheses} We define 'peak location' as the time with maximal condidial discharge, and 'peak decrease' as the average decrease in discharge between 'peak location' and end of the experiment:
\begin{equation*}
\text{peak location}_j = \arg\max \theta_j, \quad \text{peak decrease}_j = \frac{\max (\theta_j) - \theta_j(120 h)}{120h - \text{peak location}_j}
\end{equation*}
Note that this is defined on log-scale, so peak decrease should be interpreted as a relative decrease of conidial discharge.

One can qualitatively assess the hypotheses without strict definitions, but in order to do statistical inference, a mathematical definition is needed. 
We remark that here we consider population means; temporal variation may also affect peak location for individual fungi.

\section{Results}
\begin{figure}[!htb]
	\includegraphics[width=\textwidth]{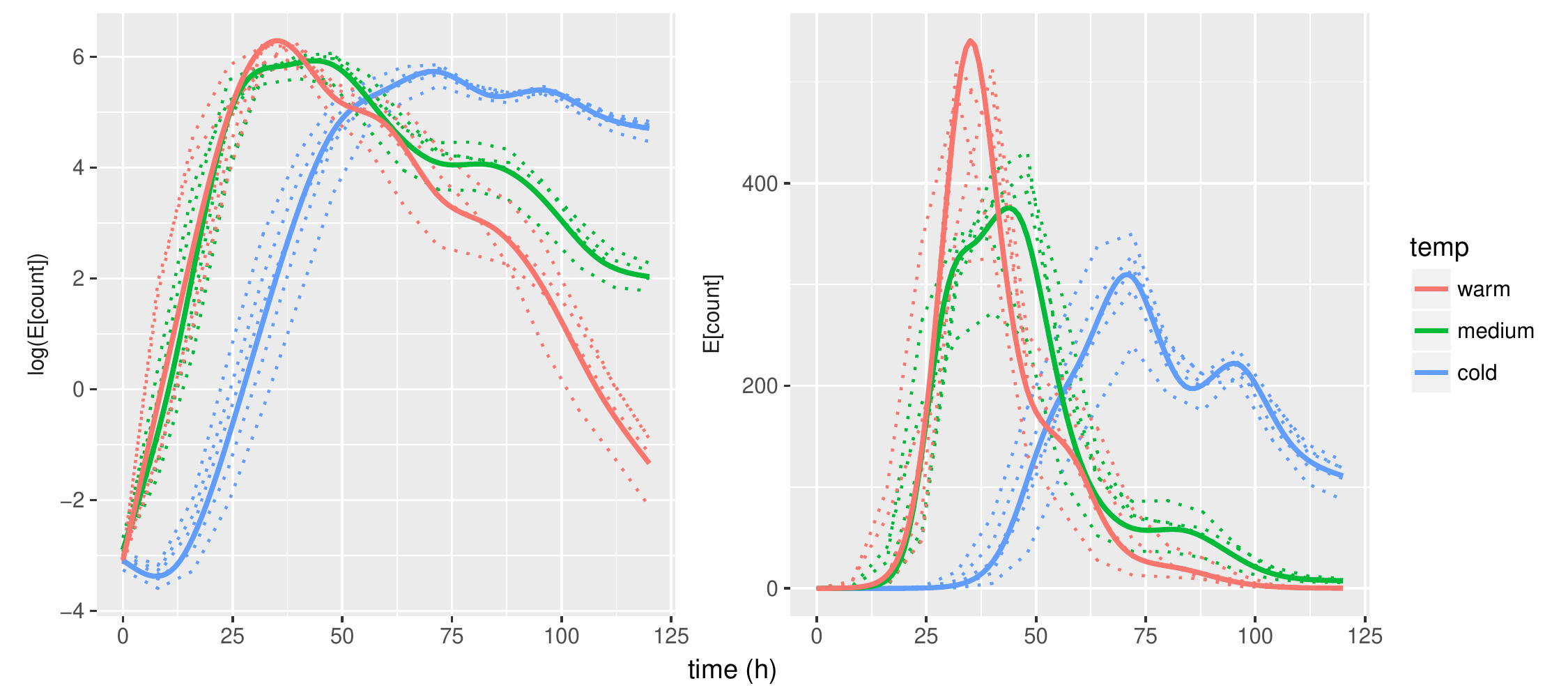}
	\caption{Predicted trajectories for $u$ (dashed lines). Left is on model scale, right is exp-transformed (same scale as the observations). Thick lines indicate estimated population means.}  \label{fig-mean-curve}
\end{figure}

Predicted mean trajectories for $u$, evaluated at observed time points, along with population means are displayed in Figure \ref{fig-mean-curve}. 
We observe a slightly odd behaviour around $t = 0$. This is an artifice; most observations around $t = 0$ are zero. When  the predicted values of $u$ are exp-transformed, these are mapped into almost-zero values. 
In concordance with our hypothesis, the
 three population means are clearly separated and fit well into what we expected: 
$\theta_{warm}$ peaks first and has the highest peak; $\theta_{middle}$ is in-between and  $\theta_{cold}$ peaks latest and has a smaller and more slowly decreasing peak.  

\begin{figure}[!htb]	
	\includegraphics[width= \textwidth]{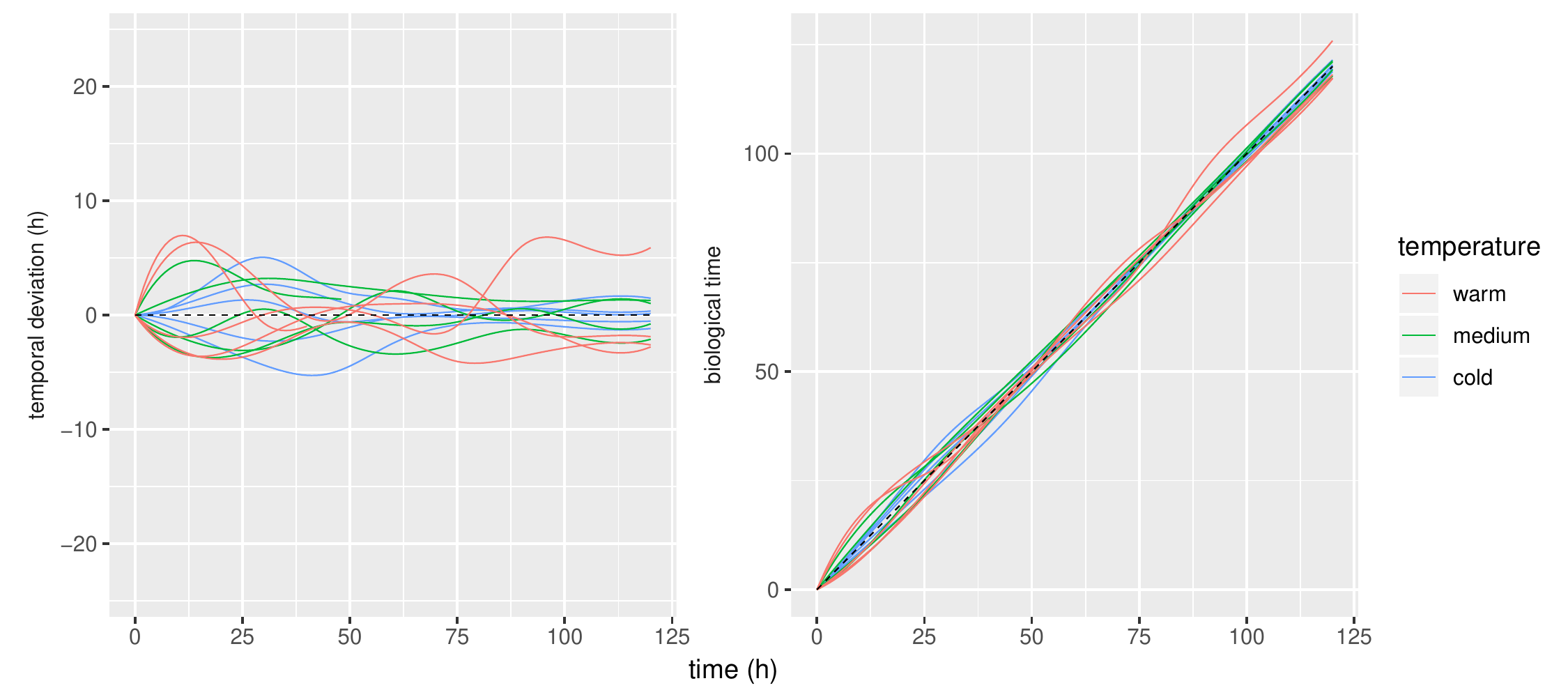}
	\caption{Predicted warp functions. Left panel: deviations from the identity. Right panel: resulting warping functions. Black line indicates the identity, ie. no temporal deviation.}  \label{fig-warp-f}
\end{figure}

Predicted warping functions are displayed in \figmark \ref{fig-warp-f}. The scale pa\-ra\-me\-ter for the warp covariance was estimated to be 0.026; this corresponds to a standard deviation of around 3.1 hours on temporal displacement, or a 95\% prediction interval of roughly 6 hours. 

The results in  \figmark \ref{fig-warp-f} are closely connected with those in \figmark \ref{fig-mean-curve}: a vertical change in \figmark  \ref{fig-warp-f} corresponds to a horisontal change in \figmark \ref{fig-mean-curve}. One may interpret the trajectories in Figures \ref{fig-mean-curve} and \ref{fig-warp-f} as \emph{smoothing} of the data: \figmark \ref{fig-data-sum} shows the raw data counts; \figmark \ref{fig-mean-curve} displays the smoothed curves, which are our predictions of the intensity of conidial discharge (the underlying biological quantity of interest) for individual fungi; and finally \figmark \ref{fig-warp-f} displays the corresponding predictions of the biological times. 

The trajectory for an individual fungus is of little interest by itself as that fungus is confined to this experiment. However, when the trajectories are viewed together, they illustrate the variation on population level allowing us to assess variation between individual fungi from the same treatment group, and also to compare this to fungi from other treatment groups.

\paragraph{Discharge of conidia above certain levels}
For practical applications it is relevant to know when the intensity of conidia discharges reaches a given level and for how long this happens. Although one conidium is enough to infect an insect \citep{yeo2001}, the chance of a conidium landing on an insect is small. Therefore we chose a range from low to very high discharge of conidia. The lowest threshold was 0.5 and  the largest threshold was 5.5 with a step size of 0.5. One step corresponds to an increase in conidia discharge of $\approx$ 65\%.
Using the results of the analysis, we simulated trajectories of $u$ from the model. For a given trajectory and threshold, we measured the first time this threshold was reached, and for how long $u$ remained above this level.

\begin{figure}[!htb]
\includegraphics[width=\textwidth]{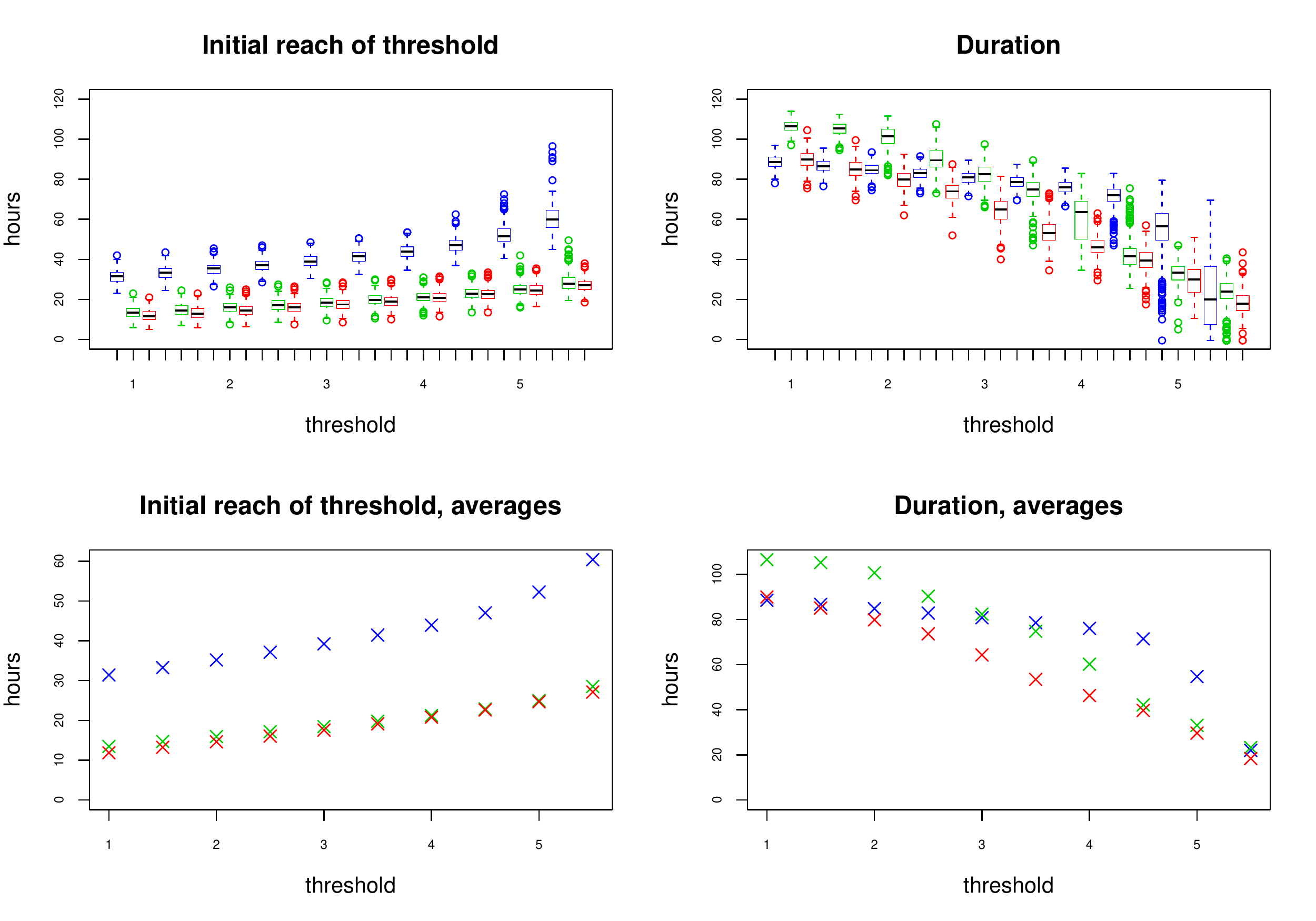}
	\caption{{Left}: First time conidial discharge intensity reaches given threshold according to the model, for different thresholds. Some trajectories did not reach given thresholds and have been omitted from the corresponding boxplots. {Right}: Duration that conidial discharge intensity is above given threshold.
	\hfill	\small \textcolor{blue}{Blue}: 12.0 \gradc, \textcolor{green}{Green}: 18.5 \gradc, \textcolor{red}{Red}: 25 \gradc.} 
	\label{fig-discharge-levels}
\end{figure}

The results are seen in \figmark \ref{fig-discharge-levels}. There are generally large variations, but we see that fungi at low temperatures are consistently slower at reaching the threshold. It should be noted that the duration is only counted until end of experiment (120 h) so the actual duration values for cold fungi could be larger when viewed over a longer time span.

\paragraph{Total conidia discharge} The total number of discharged conidia by individual fungi is displayed in Table \ref{tot-count-tabel}. Looking at the numbers, there is a decrease in total conidia count towards higher temperatures, also when discarding samples 7 and 13, which terminated discharge of conidia during the experiment.

\begin{table}[!htb]{
	\centering
	\begin{tabular}{ccc}
		cold & medium & warm \\ 
		\hline
		1575 & 2003 & 1742 \\ 
		2019 & 902* & 1764 \\ 
		1921 & 1510 & 787* \\ 
		2019 & 1991 & 1470 \\ 
		2323 & 1720 & 1769 \\ 
		\hline
	\end{tabular} \caption{Sums of discharged conidia. * indicate  fungi that terminated discharge of conidia during the experiment.} \label{tot-count-tabel}}
\end{table}

However, a one-way anova test gave a $p$-value of 0.075 (excluding samples 7 and 13), and pairwise Wilcoxon tests and a Kruskal-Wallis test gave even larger p-values. 
So while it is evident that temperature has an effect on conidia discharge as a function of time, we are not able to detect a significant effect of temperature on the total amount of conidia discharged within the first 120 h.

\subsection{Inference for population means}
\begin{figure}[!htb]

\hspace{-0.3cm}	\includegraphics[width=\textwidth]{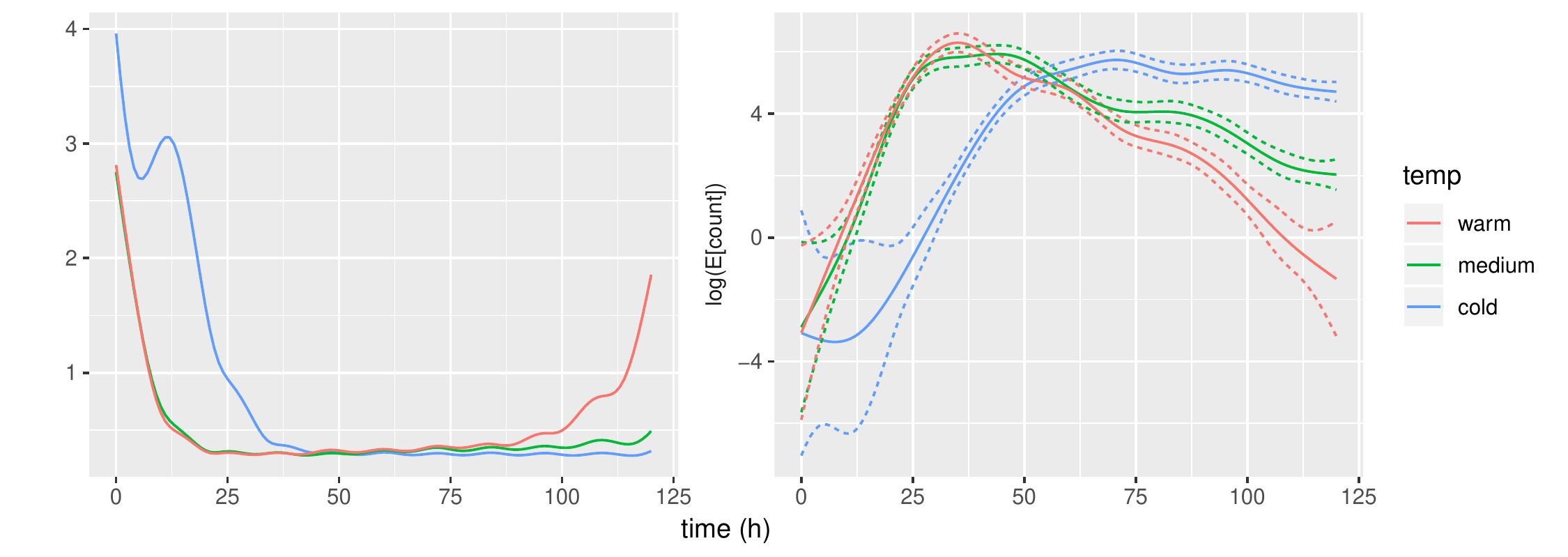}
	\caption{Left: Pointwise evaluations of $1.96 \cdot I^{-1}$, where $I$ is the information matrix. Right: Corresponding pointwise confidence intervals.} \label{confidence-plot}
\end{figure}

Following the approach outlined in the supplementary material, we estimated the information matrices for the spline coefficients, $I_{cold}, I_{medium}, I_{warm}$. 
{The information matrices themselves are of little interest, but following Berstein-von Mises theorem, the information matrix  can be used to quantify uncertainty and standard error for any given value of $\theta$, see Figure \ref{confidence-plot}.}
We have much more uncertainty for small values of $\theta$. This is as expected; small values of $\theta$ corresponds to few conidia counts and thus only little data to estimate from. The pointwise standard errors for $\theta$ in regions with large counts are around 0.20-0.25 or 20-25\% when exp-transformed.

\subsection{Peak location and decrease}
Using the standard error estimates from the previous section, we made inference on the location and decrease of peaks. This was done by simulating from the approximate distributions of the estimators. 1000 simulations were used, results are in Table \ref{peak-tabel}. As we expected,  $\theta_{cold}$ peaked late, around 70h after start, while the fungi stored at higher temperatures peaked much earlier. We observed a large and skewed 95\% confidence interval for peak location of $\theta_{medium}$, even containing the similar confidence interval for $\theta_{warm}$. 
Regarding the second element, peak decrease, we saw a roughly linear relationship between temperature and decrease. The confidence interval for $\theta_{warm}$ is broader than the other confidence intervals; this is due to the lack of data for small values of $\theta$, cf. Figure \ref{confidence-plot}.
However, all confidence intervals are clearly separated at a 95\% level, and we can firmly conclude that lower temperatures leads to a more slowly decreasing peak, with the consequence of increasing the duration of high conidial discharge.

\begin{table}[!htb]
	\centering
	\begin{tabular}{l|ccc}
		& 2.5\%  & Estimate &  97.5\% \\ 
		\hline
		cold & 66.0 & 70.7 & 73.7 \\ 
		medium & 32.7 & 43.8 & 46.6 \\ 
		warm & 33.7 & 35.1 & 36.1 \\ 
	\end{tabular} 
	\begin{tabular}{l|ccc}
		& 2.5\%  & Estimate &  97.5\% \\ 
		\hline
		cold & 1.29 & 2.10 & 2.99 \\ 
		medium & 4.14 & 5.07 & 5.98 \\ 
		warm & 6.78 & 8.94 & 11.42 \\ 
	\end{tabular} 
	\caption{Approximate 95\% confidence intervals for peak location (left) and peak decrease (right). Units are hours after start of experiment and \%/h, respectively.} \label{peak-tabel}
\end{table}

\paragraph{Credibility of hypotheses} 
By comparing the approximate distributions of the estimators, we assessed the credibility of the hypotheses stated in the data analysis section. 
This was done by pairwise comparison of estimators using $q = P( f(\hat{X}) < f(\hat{Y}))$, where $f(\hat{X})$ and $f(\hat{Y})$ were sampled independently under the posterior distributions of the parameter functions, e.g. $f(X) = peak(\theta_{cold})$ and $f(Y) = peak(\theta_{middle})$.

Identical posterior distributions of $f(\hat{X})$ and $f(\hat{Y})$ implies $q = 0.5$, 
so small or large values of $q$ are evidence against the hypothesis $f(X) = f(Y)$. 
Results are shown in Table \ref{q-tabel}. Apart from peak$(\theta_{middle})$ = peak$(\theta_{warm})$, all $q$-values are very close to one. 
As a result, our analysis very strongly supports that higher temperatures lead to faster decreasing peaks, and that a low  temperature gives a late peak in comparison to middle and high temperatures. 

\begin{table}[!htb]
	\centering
	\begin{tabular}{l|r}
		hypothesis & q \\ \hline
		 peak$(\theta_{cold})$ = peak$(\theta_{middle})$ & 1.00\\
		 peak$(\theta_{cold})$ = peak$(\theta_{warm})$  & 1.00\\
		 peak$(\theta_{middle})$ = peak$(\theta_{warm})$  & 0.108 \\
		 slope$(\theta_{cold})$ = slope$(\theta_{middle})$ & 0.9999 \\
		 slope$(\theta_{cold})$ = slope$(\theta_{warm})$ & 1.00 \\
		 slope$(\theta_{middle})$ = slope$(\theta_{warm})$ & 0.9993 \\
	\end{tabular} \caption{Pairwise comparisons of hypotheses with credibility values} \label{q-tabel}
	
\end{table}

\subsection{Robustness of statistical analysis} 
\paragraph{Leave-one-out-analysis}
To assess the uncertainty and robustness of the parameter estimates, a leave-one-out analysis was performed: One observation (or in our case, one curve) is removed from the data set, and the model is fitted to the reduced data set. This is done for all $N$ observations in turn, and the results are compared in the end. These estimates should preferably not differ by much; this is called \emph{robustness}; lack of robustness is an indication of overfitting, that is too many features or variables are included in the model. Robustness is related to \emph{generalised cross-validation}; see e.g. \cite{statlearn-book} for a reference. 

As our model is highly non-linear and consists of several layers, each with different parameters, it was of interest to study the robustness. 
\begin{table} \centering
	\begin{tabular}{c|cccccc}
		Parameter & nb-dispersion & range$_\text{amp}$ & smoothness$_\text{amp}$ &scale$_\text{amp}$  & range$_\text{warp}$ & scale$_\text{warp}$ \\ 
		\hline
		Lower bound & 4.40 & 0.314 & 2.30 & 0.0066 & 0.072 & 0.023 \\
		Estimate & 4.66 & 0.458 & 7.21 & 0.072 & 0.083 & 0.026 \\
		Upper bound & 5.21 & 0.523 & 10.0 & 0.084 & 0.691 & 0.034
	\end{tabular} \caption{Parameter estimates and leave-one-out results.  Note: An upper bound of 10 for the Mat\'ern-smoothness was used in the analysis.} \label{loo-tabel}
\end{table}
\begin{figure} \centering
	\includegraphics[width=0.7 \textwidth]{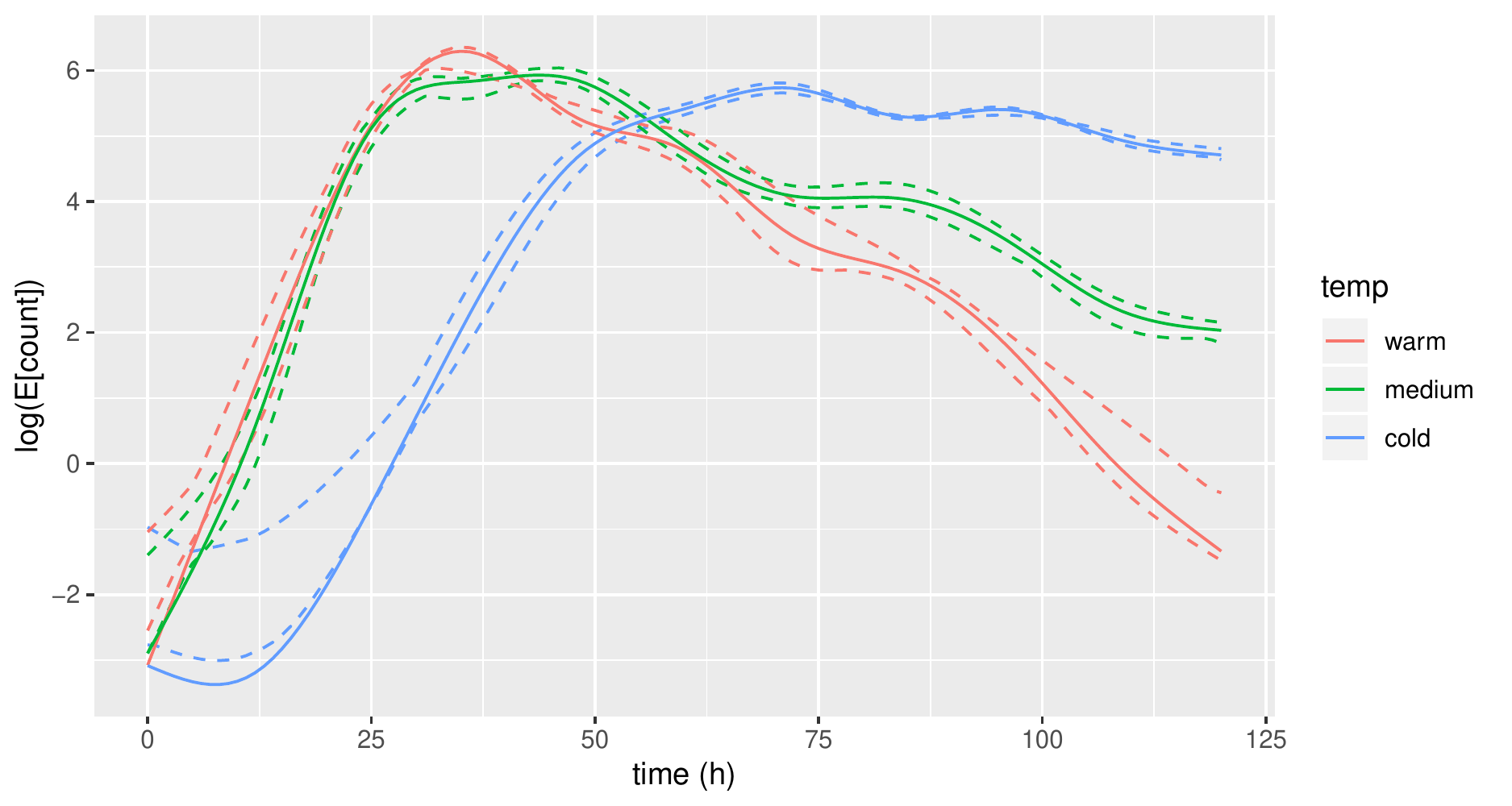}\caption{Pointwise estimates, and upper and lower bounds for leave-one-out analysis.} \label{loo-meankurver}
\end{figure}
As seen in Table~\ref{loo-tabel} we got a fairly large spread on the amplitude covariance parameters. However, this can be explained by the many kinds of variation in data; it is more relevant that the mean curves are very robust (see \figmark \ref{loo-meankurver}), as the population means are main interest of this study. The explanation behind the large spreads observed in beginning is that large negative values are mapped into almost-zero values.

\section{Discussion}

With the applied statistical methods, we were able to characterize the temporal
patterns of conidial discharge to a much better degree than previous studies, and
we characterized the variation between individual fungi at the same temperature (i.e. of the same population). With a 95\% prediction interval of roughly 6 hours, the temporal
variation is too little for changing the overall shapes, but still large enough to be
important for the analysis  and to shift the peaks for individual fungi significantly.  

Good statistical methods are essential when analysing biological systems with a temporal pattern, and allow researchers to get a better interpretation of data. Advanced statistical  methods are not always better than simple ones, but the applied methods should be able to capture all essential variations in data. The presented model accounts for all these variations, which is a major advantage to previous methods. We believe this model to be more realistic compared to other models used in similar studies. 

Examples of statistical analyses (some using the pavpop model) of other biological systems, where a model of the temporal variation was essential for the data analysis and interpretation of results, include electrophoretic spectra of cheese \citep{ronn}, growth of boys \citep{olsen2018} and hand movements \citep{RaketGrimme2016, olsen2018}. 

In this study we demonstrated the flexibility of the pavpop model by successfully fitting to a complete different kind of data: namely discrete data with many zeros, where a
Gaussian approximation would be unreasonable. With this success, there is
reason to believe that this framework would work well in applications with other
commonly used response models, for example binary response models (logistic regression).

Having several counts per measurement allowed us to look into response models. The Poisson model was invalidated, so we applied a negative binomial model instead.
This is also relevant for similar/future studies: a negative binomial distribution gives rise to larger standard errors on estimates than a corresponding Poisson model.
{Thus, if one naïvely applies a Poisson model, where a negative binomial model is correct, this increases the risk of making type I errors.}

There were some \tc{non-robustnesses} in the estimation, but given the comparatively small amount of data, this is adequate. The robustness analysis can be used to asses which parameters are identifiable in practice. Although some of the variance parameters were not well identified, the dispersion parameter, average temporal deviation and population means were found to be  robust.

We were not able to detect significant differences in total number of discharged conidia in this study. However, the fungi stored at 12 \gradc{} were still discharging many conidia at $t = 120h$, so there is good  reason to believe that there would have been significant differences if a longer time span had been used; the authors have data that supports
this, too. In a study conducted  on mycelial mats of \speccite{Pandora neoaphidis} over 168 h this could be observed: At 25 \gradc{} the mycelium mats produced less primary conidia compared to mycelium mats incubated at 10, 15 or 20 \gradc{} \citep{shah2002effects}. Aphid cadavers infected by \speccite{P. neoaphidis} discharged similar numbers of primary conidia at temperatures between 5 and 25 \gradc{} in the first 24 hours \citep{yu1995studies}.

On the other hand, we detected significant differences in peak location and shape: high temperature leads to an early peak but fast decreasing intensity of conidial discharge compared to low temperature. 
Other authors also found an earlier peak and faster decreasing intensity of conidial discharge at 25 \gradc{} compared to lower temperatures in other species of fungi, but the position of the peak and decrease of conidial discharge intensity was not statistically analysed \citep{kalsbeek2001, li2006factors}.
Our findings agree with those of \cite{li2006factors}; lower temperature leads to longer durations of conidial discharge. When the host population is large, the chance of a conidium landing on a host is larger and there is no advantage of prolonging the conidial discharge  \cite{li2006factors}.

Our work can also be seen in the perspective of disease forecasting and fungal pathogen modelling \cite{holtslag, zhouguo}, where good models for spore discharge are an important ingredient, and we believe there is an important potential (in this direction) for future research.

\paragraph{Biological control} 
There is an important plant protection perspective in this study as the fungus considered in this study has a high virulence against  insects from the genus \speccite{Cacopsylla} (Hemiptera, Psyllidae). Psyllids {harm} fruit trees by sucking from the leaves, and furthermore, the species \speccite{C. picta} can transmit plant diseases {Candidatus (Ca.) Phytoplasma pyri} to
pear trees and {Ca. Phytoplasma mali} to apple trees, causing large economic losses \cite{Strauss09}. 
There is an ongoing effort to reduce the usage of chemical pesticides for controlling psyllids in fruit and to switch to alternative control methods \cite{herrenthesis, pictakill}.

The effects of temperature on temporal pattern of conidial discharge are important in practical applications and for the potential of this species as a biocontrol agent. The most important factor is the duration of intense conidial discharge, thus we believe the biocontrol potential to be largest at cold temperatures; the effect is illustrated in \figmark \ref{confidence-plot}.
To get a better understanding of the environmental tolerance of a fungus regarding conidial discharge, experiments including fluctuating temperature, different relative humidity and light levels need to be conducted. Furthermore, the conidial discharge from insect cadavers can be measured to get a better understanding of the development of epizootics in the field. The presented statistical framework will likely be of great benefit for future data analysis of any experiments in which conidial discharge is measured over time.

\section*{Acknowledgements}
The biological experimental work was in parts supported by {the} Federal Ministry of Food, Agriculture and Consumer Protection (BMEL) in Germany, project PICTA-KILL. We thank Dr. Anant Patel (Universität Bieleveld, Germany) and Dr Jürgen Gross (Julius Kühn Institut, Dossenheim, Germany) for their support and laboratory technician Gertrud Koch for maintaining fungal cultures.

\newpage

{\centering\textbf{\Large{Supplementary material}}}

\subsection*{Statistical estimation}
		
	Direct estimation in the statistical model \eqref{u-def} and \eqref{eksp-familie} is not feasible due to the large number of latent variables. Furthermore, unlike the setup in \cite{olsen2018}, the response is not Gaussian, which further complicate estimation. One solution would be to use MCMC methods, which are generally applicable. However, we propose to use a double Laplace approximation for doing approximate maximum likelihood estimation. 
	
	This actually consists of a linearisation around the warp variables $w_n$ followed by a Laplace approximation on the discretised mean curves $\bu$; $\bu_n = \{u_n(t_{nk})\}_{k=1}^{m_n}$ for $n = 1, \dots, N$. When these approximations are done at the maximum posterior values of $(w_n, \bu_n)$, this is equivalent to a Laplace approximation jointly on $(w_n, \bu_n)$.

The main difference from the estimation procedure of  \cite{olsen2018} is the  non-trivial addition of a second layer of latent variables, $\bu$. 
		
		\paragraph{Posterior likelihood} To perform Laplace approximation, we need the mode of the joint density of observations and latent variables; this can be found by maximising the posterior likelihood for the latent variables. 
		
		The joint posterior negative log-likelihood for sample $n$ is proportional to
		\begin{equation}
		L = \big[\sum_{k = 1}^{m_n} A(u_{nk})  - u_{nk}y_{nk}\big]
		+ \frac{1}{2} (\bgamma_{w_n} - \bu_n)\trans S_n^{-1} (\bgamma_{w_n} - \bu_n) + \frac{1}{2} w_n^\top C^{-1} w_n
		\label{post-lik-formel}
		\end{equation}
		where $	\bgamma_{w_n}$ denote the vector $\{\theta_{f(n)}(v(t_{nk} , w_n)) \}_{k=1}^{m_n}$. Spline coefficients for the fixed effects are indirectly present in the posterior likelihood through $\bgamma_{w_n}$; more details follow below.
		It should be noted that under relatively mild assumptions, minimizing \eqref{post-lik-formel} for a fixed $w$ is a convex optimization problem.
		
\paragraph{Likelihood approximation} To approximate the likelihood, %
we firstly linearise around $w^0$ to approximate $p(u)$ with a Gaussian distribution and secondly we make a Laplace approximation of the joint linearised likelihood.

The linearization around   $w^0$ to approximate the likelihood for density the mean curves, $p(u_n)$, is described in detail in \citep{RaketSommerMarkussen, olsen2018}.
 The result of doing this is a Gaussian approximation of the latent $u$, ie. $\bu_n \stackrel{D}{\approx} \tilde{\bu}_n$ where $\tilde{\bu}_n \sim N(r_n, V_n)$. 
$r_n$ and $V_n$ are obtained from the Taylor approximation of $u$ around the posterior maximum $w_n^0$; for details we refer to \citep{RaketSommerMarkussen, olsen2018}.

In general, the Laplace approximation of an integral on the form $\int_{\R^d} e^{f(x)} \de x$ around the mode $x_0$ of $f$ is given by
\begin{equation}
(2 \pi)^{d/2} |-f''(x_0)|^{-1/2} e^{f(x_0)}
\end{equation}
where $|-f''(x_0)|$ is the determinant of the Hessian of $-f$, evaluated in $x_0$. This approximation is exact if $f$ is a second-order polynomial, and generally the approximation is directly related to the second-order Taylor approximation of $f$ at $x_0$. 

Up to some constants, which do not depend on the parameters, the likelihood for a single curve in the linearised model is given by the following integral, which we want to approximate:
\begin{equation}
L^\text{lin}_n \propto \inte{\R^{m_n}}{} {|{V_n}|^{-1/2} \exp \left(- \tfrac{1}{2}  (\bu_n - r_n) \trans V_n^{-1}  (\bu_n - r_n) + 
	\sum_{k=1}^{m_n} y_{nk} \bu^0_{nk} - A(u_{nk}) \right) }{\bu_n}
\end{equation}

Assuming $\bu_n^0$ to be the maximum of the posterior likelihood \eqref{post-lik-formel}, one can show that the negative logarithm of the Laplace approximation around $(\bu_n^0, w_n^0)$ is given by:
\begin{equation*}
1/2 \log |\tilde{\Sigma}_n|  +  \sum_{k=1}^{m_n} (  A(u^0_{nk}) - y_{nk} \bu^0_{nk}) + p(\bu^0_n)
\end{equation*}
where  $\tilde{\Sigma}_n= V_n^{-1} + 2 diag(A''(\bu_n^0))$ and $p(\cdot)$ is the negative log-density for a $N(r_n^0, V_n)$-distribution. By assumption, $A''(u^0_{nk}) > 0$, so $|\tilde{\Sigma}_n| > |V_n|^{-1}$.

The total log-likelihood for all observations is then approximated by
\begin{equation}
\sum_{n=1}^N  \left[\log | \tilde{\Sigma}_n | +  \log|V_n| +  (\bu_n^0 - r_n) \trans V_n^{-1}  (\bu_n^0 - r_n) + 2
\sum_{k=1}^{m_n} (  A(u^0_{nk}) - y_{nk} \bu^0_{nk}) \right] \label{loglik-tot} 
\end{equation}	

\paragraph{Inference} We propose to use alternating steps of (a) estimating spline coefficients for the fixed effects and predicting the most likely warps and mean curves by minimizing the posterior log-likelihood \eqref{post-lik-formel} and  (b) estimating variance parameters from minimizing the approximated log-likelihood \eqref{loglik-tot}.

\subsection*{Fixed effects and phase variation} \label{fix-phase}

Fixed effects are modelled using a spline basis that is assumed to be continuously differentiable, e.g. a Fourier basis or B-spline bases. 
A typical choice for non-periodic data would be B-splines; we have used natural cubic splines in the data application. 
Fixed effects are estimated using the posterior likelihood \eqref{post-lik-formel}. For a fixed value of $w_n$,  $\bgamma_{w_n}$ is a linear function of the spline coefficients, and thus the optimal value can be found using standard linear algebra tools.

Phase variation is modelled by random warping functions $v_n = v( \cdot, w_n): [0,1] \pil D$, parametrized by independent zero-mean Gaussian variables $w_n \in \R^{m_w}$. $v: [0,1] \times R^{m_w} \pil D$ is a suitable spline interpolation function, such that $v(\cdot , 0)$ is the identity on $[0,1]$. 

The latent trajectories $v_n$ are modelled as deviations from the identity function at pre-specified time points $(t_k)_{k=1}^{m_w}$, subject to a Hyman filtered, cubic spline interpolation for insuring monotonicity, $v_n(t_k) \approx t_k + w_{nk}$. A more detailed discussion of modelling phase variation using increasing spline functions can be found in \cite{olsen2018}.

\paragraph{Uncertainty for fixed effects}
As our model is highly non-linear, we cannot expect to find closed-form expressions for the uncertainty of the parameter estimates. Furthermore, the latent variables complicate assessment of uncertainty as these are uncertain themselves.

A standard quantifier for assessing uncertainty in statistical models is the \emph{information matrix}, which  can be approximated by the second-order derivative of the log-likelihood at the MLE.
However, directly using \eqref{loglik-tot} would underestimate the information, as \eqref{loglik-tot} depend on the optimal value of the posterior likelihood \eqref{post-lik-formel}, which itself is a function of parameters. 

Let $c_j$ denote the spline coefficients which determine the population mean $\theta_j$ for treatment group $j$. $c_j$ is determined from the posterior likelihood $L = L(c, u, w)$, given in Equation \ref{post-lik-formel}. As $u$ and $w$ are latent, it would be wrong to use the second derivative of $L$ for the information matrix; instead we use the second derivative of $f(c) = L(c, u(c), w(c))$, where $u$ and $w$ are viewed as functions that map $c$ into the max-posteriors of $u$ and $w$ given $c$. 

This will more correctly ensure that the uncertainty of $u$ and $w$ is taken into account when estimating the  {information matrix}. Furthermore, positive definiteness of $L''$ will imply positive definiteness of $f''$.

\subsection*{Response models}
In the application presented in this paper we assume that $(y|u)$ follows a negative
binomial distribution. There are various choices of response models,
a list of important ones are stated below. Note that not all exponential families fits naturally with our methodology; $y|u$ must be well-defined for all $u \in \R$.
	
\paragraph{Binary response:}

For binary responses, the sample space is $\mathcal{Y} = \{0, 1\}$. If we define $p := P(Y = 1 | \eta)$, and set  $A(\eta) = \log(1 + e^\eta)$, we get that $\eta = \log(p)- \log(1-p)$, the canonical link function for regression models with binomial response.
	
\paragraph{Poisson model:} For the Poisson model we have $Y \in \N_0$ where $A(\eta) = e^{\eta}$.
The conditional mean satisfies $\E[Y | \eta] = e^{\eta}$, and by inverting this relation we get $\eta = \log \E[y | \eta]$, the canonical link function. 
	
\paragraph{Negative binomial model:} Negative binomial distributions are often viewed as overdispersed versions of Poisson models. 
	Let the rate parameter $r > 0$ be given such that $V[Y | \eta] =  \E[Y | \eta] + \E[Y | \eta]^2/r$; the limit $ r \pil \infty$ corresponds to the Poisson model.
	
	We get $A(\eta, y) = (r + y) \log(1 + \frac{e^\eta}{r})$ and $A''(\eta; y) = (y+r) \frac{r e^\eta}{(r+ e^\eta)^2}$. Unlike the Poisson and binomial models, the link function $A$ depends on $y$, but it is easily seen that $A(\eta, y) $ approximates  $e^{\eta}$ in the limit $r \pil \infty$. 
	
\paragraph{Normal distribution with known variance $\sigma^2$:} For normal distributions we have $Y \in \R$. By setting $\tilde{Y} = Y / \sigma^2$, then $A(\eta) = \eta^2/2\sigma^2$, $\E[\tilde{Y} | \eta] = \eta/\sigma^2 $, and $\E[Y | \eta] = \eta$. This is the most basic response model, and the one used in \cite{RaketSommerMarkussen}. \cite{RaketSommerMarkussen} use a different formulation and also treats  $\sigma^2$ as an unknown parameter. The Laplace approximation becomes exact when using normal distributions, simplifying estimation to become the approach used in \cite{olsen2018}.
	
\subsection*{Matern covariance function}	
The Matérn covariance function is commonly used in functional data analysis and spatial statistics. It is given by
\begin{equation}
f_{\sigma, \alpha, \kappa}(s,t) = \sigma^2 \frac{2^{1-\alpha}}{\Gamma(\alpha)} \left(\frac{\alpha |s-t|}{\kappa} \right)^{\alpha}
K_\alpha\left(\frac{\alpha |s-t|}{\kappa} \right), \quad s,t \in \R
\end{equation}
where $K_\alpha$ is the modified Bessel function of the second kind. Here $\sigma$ is the scale parameter, $\alpha$ is the smoothness parameter and $\kappa$ is the range parameter.
	
\end{document}